\documentclass[11pt,a4paper]{article}

\usepackage{amssymb}
\usepackage[dvips]{graphicx}
\usepackage{bm}

\begin{document}

\title{Gauge dependence of the one-loop divergences \break in $6D$, ${\cal N} = (1,0)$ abelian theory}

\author{I.L. Buchbinder\footnote{joseph@tspu.edu.ru}\\
{\small{\em Department of Theoretical Physics, Tomsk State Pedagogical
University,}}\\
{\small{\em 634061, Tomsk,  Russia}} \\
{\small{\em National Research Tomsk State University, 634050, Tomsk, Russia}},\\
{\small{\em Bogoliubov Laboratory of Theoretical Physics, JINR, 141980 Dubna, Moscow region,
Russia}},\\
\\
E.A. Ivanov\footnote{eivanov@theor.jinr.ru}\\
{\small{\em Bogoliubov Laboratory of Theoretical Physics, JINR, 141980 Dubna, Moscow region,
Russia}},\\
\\
B.S. Merzlikin\footnote{merzlikin@tspu.edu.ru}\\
{\small{\em Department of Theoretical Physics, Tomsk State Pedagogical
University}},\\
{\small{\em 634061, Tomsk,  Russia}}, \\
{\small{\em Division of Experimental Physics}},\\
{\small{\em \it Tomsk Polytechnic University, 634050, Tomsk, Russia}},\\
{\small{\em Bogoliubov Laboratory of Theoretical Physics, JINR, 141980 Dubna, Moscow region,
Russia}},\\
\\
K.V. Stepanyantz\footnote{stepan@m9com.ru}\\ {\small{\em Moscow State University}},\\
{\small{\em Faculty of Physics, Department of Theoretical Physics}},\\
{\small{\em 119991, Moscow, Russia}},\\
{\small{\em Bogoliubov Laboratory of Theoretical Physics, JINR, 141980 Dubna, Moscow region,
Russia}} }

\date{}

\maketitle

\begin{abstract}
We study the gauge dependence of the one-loop effective action for the abelian $6D$, ${\cal N}=(1,0)$ supersymmetric gauge theory formulated in harmonic superspace. We introduce the superfield $\xi$-gauge, construct the corresponding gauge superfield propagator, and calculate the one-loop two-and three-point Green functions with two external hypermultiplet legs. We demonstrate that in the general $\xi$-gauge the two-point Green function of the hypermultiplet is divergent, as opposed to the Feynman gauge $\xi =1$. The three-point Green function with two external hypermultiplet legs and one leg of the gauge superfield is also divergent. We verified that the Green functions considered satisfy the Ward identity formulated in ${\cal N}=(1,0)$ harmonic superspace and that their gauge dependence vanishes on shell. Using the result for the two- and three-point Green functions and arguments based on the gauge invariance, we present the complete divergent part of the one-loop effective action in the general $\xi$-gauge.
\end{abstract}

\unitlength=1cm

\section{Introduction}
\hspace*{\parindent}

Gauge theories with extended supersymmetries in higher
dimensions attract a considerable attention for a long time
\cite{Howe:1983jm,Howe:2002ui,Bossard:2009sy,Bossard:2009mn,Fradkin:1982kf,Marcus:1983bd,Smilga:2016dpe,Bork:2015zaa}.
On the one hand, such theories are non-renormalizable due to the
dimensionful coupling constant (see, e.g.,
\cite{Gates:1983nr,Buchbinder:1998qv}). On the other hand, one can
expect an improvement of the ultraviolet behavior due to the extended
supersymmetry.  It is very interesting to check this conjecture on the explicit examples of
higher-dimensional supersymmetric theories. To be more realistic, one can expect that the full canceling of divergences is presumably possible only in the lowest loops even
in the maximally extended theories (see, e.g., \cite{Marcus:1984ei}).
The problem reveals clear analogies with the most interesting case of gravity. However, the analysis in supersymmetric gauge theories is much simpler.

In order to fully display the underlying properties of theories with some
symmetries it is highly desirable to be aware of the regularization and
quantization schemes which do not break these symmetries. In the
case of extended supersymmetries these purposes can be achieved within the
harmonic superspace approach
\cite{Galperin:1984av,Galperin:1985bj,Galperin:1985va,Galperin:2001uw,Buchbinder:2001wy,Buchbinder:2016wng}.
For $6D$ supersymmetric gauge theories (which will be  the subject of  the present paper) this
formalism
\cite{Howe:1983fr,Howe:1985ar,Zupnik:1986da,Ivanov:2005qf,Ivanov:2005kz,Buchbinder:2014sna}
ensures manifest ${\cal N}=(1,0)$ supersymmetry. With the use
of the background field method in  harmonic superspace
\cite{Buchbinder:2001wy,Buchbinder:1997ya}, gauge symmetry can also be
made manifest. For these reasons the harmonic superspace formalism seems to be
most suitable for quantum calculations in $6D$
supersymmetric theories (note that $6D\,,$ ${\cal N}=(1,0)$ theories
are in general anomalous, see, e.g.,
\cite{Townsend:1983ana,Smilga:2006ax,Kuzenko:2015xiz,Kuzenko:2017xgh}).

Recently, some explicit calculations based on the harmonic superspace method were
done for ${\cal N}=(1,0)$ and ${\cal N}=(1,1)$ gauge theories
\cite{Buchbinder:2016gmc,Buchbinder:2016url,Buchbinder:2017ozh,Buchbinder:2017gbs,Buchbinder:2017xjb},
following the general pattern of Ref. \cite{Bossard:2009mn}.
These calculations were basically performed in the Feynman gauge $\xi=1$, which
ensures the simplest form of the propagator of the gauge superfield. This
considerably simplifies the calculation of quantum corrections.
However, the gauge dependence of the results obtained by the harmonic superspace technique has not been
yet analyzed.  Meanwhile, the calculations in the non-minimal
gauges are frequently rather useful as compared to those in the Feynman gauge, because they  are capable to make manifest
divergences in the lower loops. For example, for ${\cal N}=1$ supersymmetric gauge theories in the one-loop approximation ghosts
are not renormalized in the Feynman gauge, while divergences appear
for $\xi\ne 1$ \cite{Aleshin:2016yvj}. For calculations in
higher orders, the knowledge of gauge dependence in the lower-order approximations is
also essential, see, e.g., \cite{Kazantsev:2018kjx}. These are the reasons why
a vast literature is devoted to calculations in non-minimal gauges. As a characteristic example, let us mention  a
recent paper \cite{Chetyrkin:2017mwp}.

In the present paper we consider the simplest $6D\,,$ ${\cal N}=(1,0)$
supersymmetric gauge theory, namely, ${\cal N}=(1,0)$ supersymmetric
electrodynamics, and investigate the structure of the gauge-dependent
contributions to the effective action by the harmonic
superspace technique. In particular, we demonstrate that (unlike the
case of the Feynman gauge considered, e.g., in
\cite{Buchbinder:2016gmc}) the two-point Green function of
hypermultiplets is divergent already at the one-loop level.
The gauge-dependent divergences are also present in the
gauge multiplet - hypermultiplet Green functions. In this paper we explicitly calculate
 the one-loop three-point Green function and find its
divergent part. Moreover, we derive the Ward identity in the harmonic
superspace and verify that the Green functions obtained by
calculating harmonic supergraphs satisfy this identity, as expected. This
result is a non-trivial verification of the correctness of our calculations.
One more test, which has also been done in this paper, is the demonstration of the property that the
gauge dependence of the effective action vanishes on shell (this is
a consequence of the general theorem, see Refs.
\cite{DeWitt:1965jb,Boulware:1980av,Voronov:1981rd,Voronov:1982ph,Voronov:1982ur,Lavrov:1986hr}).
Using the results for the two- and three-point Green functions, we
also restore the complete result for the one-loop divergences, based on the
gauge invariance of the theory under consideration.

The paper is organized as follows: In Sect. \ref{Section_Electrodynamics} we recall some basic points of the formulation of $6D\,,$ ${\cal N}=(1,0)$
supersymmetric electrodynamics in harmonic superspace. We present the superfield action for this theory, write down the Ward identity,
and formulate the harmonic superspace Feynman rules. In particular, we construct the propagator of the gauge superfield in the non-minimal
gauges which are analogs of the $\xi$-gauges in the usual electrodynamics. In Sect. \ref{Section_Divergences},
using these Feynman rules, we investigate the gauge dependence of the one-loop two-point Green functions of the gauge superfield
and the hypermultiplet. We also calculate the one-loop three-point gauge superfield - hypermultiplet Green function. Checking the Ward identities
for these Green functions is the subject of Sect. \ref{Section_Ward}. The vanishing of the gauge dependence on shell in the approximation we are considering
is demonstrated in Sect. \ref{Section_Shell}. The total divergent part of the one-loop effective action (which is an infinite series in $V^{++}$)
is constructed in Sect. \ref{Section_Total_Divergences}, by invoking the arguments based on the gauge invariance. Also we verify that
the gauge dependence of the expression obtained vanishes on shell. Some technical details are collected in two Appendices.

\section{Harmonic superspace formulation of $6D\,,$ ${\cal N}=(1,0)$ electrodynamics}
\label{Section_Electrodynamics}

\subsection{The harmonic superspace action}
\hspace*{\parindent}

The harmonic superspace is very convenient for formulating $6D$\,, ${\cal N}=(1,0)$ supersymmetric theories,
because it ensures manifest supersymmetry at all steps of quantum calculations.
It is parametrized by the coordinate set $(x^M,\theta^{ai}, u_i^\pm)$ which will be referred to as the central basis. Here $x^M$ with $M=0,\ldots 5$ are the usual
coordinates of the six-dimensional Minkowski space. The Grassmann anticommuting coordinates $\theta^{ai}$
with $a=1,\ldots 4$ and $i=1,2$ form a left-handed $6D$ spinor. The harmonic variables $u_i^\pm$ satisfy
the condition $u^{+i} u_i^- = 1$, with $u_i^- = (u^{+i})^*$. The analytic basis of the harmonic superspace is parametrized
by the coordinates

\begin{equation}
x^M_A = x^M + \frac{i}{2}\theta^{-}\gamma^M \theta^+;\qquad \theta^{\pm a} = u^\pm_i \theta^{ai}, \quad u^{\pm}_i\,,
\end{equation}

\noindent
where $\gamma^M$ are $6D$ $\gamma$-matrices. The coordinate subset $(x^M_A, \theta^{+ a}, u^\pm_i)$ parametrizes the analytic harmonic subspace which
is closed on its own under $6D\,, {\cal N}=(1,0)$ supersymmetry transformations.

It is convenient to define the spinor covariant derivatives

\begin{equation}
D^+_a = u^{+}_i D_{a}^i;\qquad D^-_a = u^{-}_i D_{a}^i,
\end{equation}

\noindent
such that $\{D^+_a, D^{-}_b\} = i(\gamma^M)_{ab}\partial_M$, and to introduce the notation

\begin{equation}
(D^+)^4 = -\frac{1}{24}\varepsilon^{abcd} D_a^+ D_b^+ D_c^+ D_d^+.
\end{equation}

\noindent
Also we will need the harmonics derivatives in the central basis

\begin{equation}
D^{++} = u^{+i} \frac{\partial}{\partial u^{-i}};\qquad D^{--} = u^{-i} \frac{\partial}{\partial u^{+i}};\qquad
D^0 = u^{+i} \frac{\partial}{\partial u^{+i}} - u^{-i} \frac{\partial}{\partial u^{-i}}.
\end{equation}

\noindent
They satisfy the commutation relations of the $SU(2)$ algebra. The analytic basis form of these derivatives can be easily found and is given, e.g., in \cite{Bossard:2015dva}.

For constructing the ${\cal N}=(1,0)$ invariants we need the invariant superspace integration measures:

\begin{eqnarray}\label{Measure_Total}
&& \int d^{14}z = \int d^6x\,d^8\theta;\qquad \int d\zeta^{(-4)} = \int d^6x\, d^4\theta^+;\\
\label{Measure_Analytic}
&& \int d^6x\,d^8\theta = \int d^6x\,d^4\theta^{+} (D^+)^4.
\end{eqnarray}

In this paper we consider ${\cal N}=(1,0)$ supersymmetric electrodynamics, which is a particular abelian  case of ${\cal N}=(1,0)$ supersymmetric
Yang--Mills theory with hypermultiplets. The harmonic superspace form of the action of $6D$, ${\cal N}=(1,0)$ supersymmetric Yang--Mills
theory was pioneered in Ref. \cite{Zupnik:1986da}. As opposed to the analogous $4D$, ${\cal N}=2$ construction, the gauge theory coupling constant $f_0$ in $6D$ has the dimension $m^{-1}\,$.
In the harmonic superspace approach the gauge superfield $V^{++}(z,u)$ satisfies the analyticity condition

\begin{equation}
D^+_a V^{++} = 0
\end{equation}

\noindent
and is real with respect  to  the special conjugation denoted by\ \ $\widetilde{}$\ , i.e. $\widetilde{V^{++}} = V^{++}$.
The hypermultiplets are described by the analytic superfield $q^+$ and its\ \ $\widetilde{}$\ -conjugate $\widetilde q^+$.

Like in the non-supersymmetric case, the action of ${\cal N}=(1,0)$ electrodynamics is quadratic in the gauge superfield. It can be written as

\begin{equation}\label{Electodynamics_Action}
S = \frac{1}{4f_0^2} \int d^{14}z\,\frac{du_1 du_2}{(u_1^+ u_2^+)^2} V^{++}(z,u_1) V^{++}(z,u_2) - \int d\zeta^{(-4)} du\,\widetilde q^+ \nabla^{++} q^+.
\end{equation}

\noindent
where

\begin{equation}\label{Gauge_Covariant_Derivative}
\nabla^{++} = D^{++} + i V^{++}
\end{equation}

\noindent
and $D^{++}$ is taken in the analytic basis. The gauge transformations has the form

\begin{equation}\label{Gauge_Transformations}
V^{++} \to  V^{++}  - D^{++} \lambda; \qquad  q^+ \to  e^{i\lambda} q^+; \qquad  \widetilde q^+ \to  e^{-i\lambda} \widetilde q^+,
\end{equation}

\noindent where $\lambda$ is an analytic superfield parameter which is real with respect to the\ \ $\widetilde{}$\ -conjugation.

It is useful to introduce the non-analytic superfield

\begin{equation}\label{V--}
V^{--}(z,u) = \int du_1\,\frac{V^{++}(z,u_1)}{(u^+ u_1^+)^2}.
\end{equation}

\noindent
It satisfies the conditions $D^{++} V^{--} = D^{--} V^{++}$ and transforms as

\begin{equation}
V^{--} \to V^{--}  - D^{--}\lambda
\end{equation}

\noindent
under the gauge transformations. Starting from this superfield, it is possible to construct
the analytic superfield $F^{++} = (D^+)^4 V^{--}$, which is gauge invariant in the abelian case.

For further use, we also define the non-analytic superfield $q^-$ as a solution of the equation

\begin{equation}\label{Q-_Definition}
q^+ = \nabla^{++} q^- = (D^{++} + i V^{++}) q^-.
\end{equation}

\noindent
From this definition one can  derive that the gauge transformations act on $q^-$ as

\begin{equation}
q^- \to e^{i\lambda} q^-.
\end{equation}

\noindent
In the explicit form the solution of Eq. (\ref{Q-_Definition}) can be expressed as the series

\begin{eqnarray}\label{Q-_Expansion}
&&\hspace*{-8mm} q^- = \int \frac{du_1}{(u^+ u_1^+)} q_1^+ -i \int  \frac{du_1\,du_2}{(u^+ u_1^+)(u_1^+ u_2^+)} V^{++}_1 q_2^+ - \int  \frac{du_1\,du_2\,du_3}{(u^+ u_1^+)(u_1^+ u_2^+)(u_2^+ u_3^+)} V^{++}_1 V^{++}_2 q_3^+ + \ldots\nonumber\\
&&\hspace*{-8mm}  =(-i)^{n-1} \sum\limits_{n=1}^\infty \int du_1 \ldots du_n\, \frac{V^{++}_1 \ldots V^{++}_{n-1}}{(u^+ u_1^+) \ldots (u_{n-1}^+ u_n^+)} q_n^+,
\end{eqnarray}

\noindent
where subscripts numerate the harmonic ``points''.

For quantizing the theory (\ref{Electodynamics_Action}) it is necessary to fix the gauge. This can be done by adding the gauge-fixing term to the action,

\begin{equation}\label{Gauge_Fixing_Term}
S_{\mbox{\scriptsize gf}} = - \frac{1}{4f_0^2\xi_0} \int d^{14}z\, du_1 du_2 \frac{(u_1^- u_2^-)}{(u_1^+ u_2^+)^3} D_1^{++} V^{++}(z,u_1) D_2^{++} V^{++}(z,u_2),
\end{equation}

\noindent
where $\xi_0$ is the bare gauge-fixing parameter. This term corresponds to the $\xi$-gauge in the usual electrodynamics. In particular, the Feynman gauge amounts to the
choice  $\xi_0=1$. In the abelian case we are considering it is not necessary to introduce the ghosts superfields. Therefore, the generating functional for our theory can be written as

\begin{equation}\label{Generating_Functional}
Z = \int DV^{++}\,D\widetilde q^+\, Dq^+\, \exp\Big\{i(S+S_{\mbox{\scriptsize gf}} + S_{\mbox{\scriptsize sources}})\Big\},
\end{equation}

\noindent
where $S_{\mbox{\scriptsize sources}}$ is a sum of the source terms,

\begin{equation}\label{Sources}
\int d\zeta^{(-4)}\,du\, \Big[V^{++} J^{(+2)} + j^{(+3)} q^+ + \widetilde j^{(+3)} \widetilde q^+\Big].
\end{equation}

\noindent
Here $J^{(+2)}$ is the analytic source for the gauge superfield, while $j^{(+3)}$ and $\widetilde j^{(+3)}$ denote sources for the hypermultiplet superfields.
The effective action is constructed from the generating functional for the connected Green functions $W = -i\ln Z$ by making the Legendre transformation,

\begin{equation}\label{Gamma_Definition}
\Gamma = W - S_{\mbox{\scriptsize sources}},
\end{equation}

\noindent
where it is necessary to express the sources in terms of the fields with the help of the equations

\begin{equation}
V^{++} = \frac{\delta W}{\delta J^{(+2)}};\qquad q^+ = \frac{\delta W}{\delta j^{(+3)}};\qquad \widetilde q^+ = \frac{\delta W}{\delta \widetilde j^{(+3)}}.
\end{equation}

\subsection{Ward identity}
\hspace*{\parindent}

In the abelian gauge theory  at the quantum level the gauge invariance is encoded in the Ward identity \cite{Ward:1950xp},
which is a particular case of the Slavnov--Taylor identities \cite{Taylor:1971ff,Slavnov:1972fg}. The harmonic superspace analog of this identity
can be formulated,  using the standard technique. For this purpose we make the  transformation (\ref{Gauge_Transformations})
in the generating functional (\ref{Generating_Functional}) which evidently remains invariant. Taking into account
that the classical action is gauge invariant, in the lowest order in $\lambda$ we obtain

\begin{eqnarray}\label{Original_Ward}
0 =  \Big\langle \int d\zeta^{(-4)}\,du\, \Big[ - \frac{\delta S_{\mbox{\scriptsize gf}}}{\delta V^{++}} D^{++}\lambda
- J^{(+2)} D^{++}\lambda + i j^{(+3)} \lambda q^+ - i \widetilde j^{(+3)} \lambda \widetilde q^+\Big]\Big\rangle,
\end{eqnarray}

\noindent
where we used the notation

\begin{equation}
\Big\langle A(V^{++}, q^+, \widetilde q^+)\Big\rangle = \frac{1}{Z} \int DV^{++}\,D\widetilde q^+\, Dq^+\, A(V^{++}, q^+, \widetilde q^+) \exp\Big\{i(S+S_{\mbox{\scriptsize gf}} + S_{\mbox{\scriptsize sources}})\Big\}.
\end{equation}

\noindent
Integrating in Eq. (\ref{Original_Ward}) by parts with respect to  the derivatives $D^{++}$, using an arbitrariness of $\lambda$, and expressing the result in terms of superfields,
we obtain

\begin{eqnarray}
0 =  D^{++} \frac{\delta S_{\mbox{\scriptsize gf}}}{\delta V^{++}}
-D^{++}\frac{\delta\Gamma}{\delta V^{++}} - i q^+ \frac{\delta\Gamma}{\delta q^+} + i \widetilde q^+ \frac{\delta\Gamma}{\delta\widetilde q^+},
\end{eqnarray}

\noindent
where $\Gamma$ is the effective action defined by Eq. (\ref{Gamma_Definition}), and we also took into account that the gauge-fixing term is quadratic in the gauge superfield. Introducing

\begin{equation}
\Delta\Gamma = \Gamma - S_{\mbox{\scriptsize gf}},
\end{equation}

\noindent
the Ward identity  can be written in a more compact form,

\begin{eqnarray}\label{Generating_Ward_Identity}
D^{++}\frac{\delta\Delta\Gamma}{\delta V^{++}} = - i q^+ \frac{\delta\Delta\Gamma}{\delta q^+} + i \widetilde q^+ \frac{\delta\Delta\Gamma}{\delta\widetilde q^+}.
\end{eqnarray}

\noindent
It is important that this equation is valid for arbitrary non-zero values of the involved superfields. Differentiating Eq. (\ref{Generating_Ward_Identity})
with respect to various superfields we derive an infinite set of identities relating the longitudinal part of the $(n+1)$-point Green functions to the $n$-point Green functions.
For example, differentiating with respect to $V^{++}_2$ and setting all fields equal to zero at the end,
we obtain that quantum corrections to the two-point Green function of the gauge superfield are transversal,

\begin{equation}\label{Transversality}
D^{++}_1 \frac{\delta^2\Delta\Gamma}{\delta V^{++}_1 \delta V^{++}_2} = 0.
\end{equation}

\noindent
Differentiating Eq. (\ref{Generating_Ward_Identity}) with respect to $q^+_2$ and $\widetilde q^+_3$ and setting the fields equal to zero  at the end give an analog
of the usual Ward identity relating three- and two-point Green functions:

\begin{eqnarray}\label{Ward_Identity}
&& D^{++}_1 \frac{\delta^3\Delta\Gamma}{\delta V^{++}_1 \delta q^+_2 \delta\widetilde q^+_3} = - i (D_1^+)^4 \delta^{14}(z_1-z_2) \delta^{(-3,3)}(u_1,u_2) \frac{\delta^2\Delta\Gamma}{\delta q^+_1 \delta\widetilde q^+_3}\nonumber\\
&&\qquad\qquad\qquad\qquad\qquad\qquad\qquad + i (D^+_1)^4\delta^{14}(z_1-z_3)\delta^{(-3,3)}(u_1,u_3) \frac{\delta^2\Delta\Gamma}{\delta q^+_2 \delta\widetilde q^+_1}.\qquad
\end{eqnarray}

\noindent
When deriving this equation, we have taken into account the property implied by the  Grassmann analyticity

\begin{equation}\label{Analytic_Derivative}
\frac{\delta q^+_1}{\delta q^+_2} = (D^+_1)^4 \delta^{14}(z_1-z_2) \delta^{(-3,3)}(u_1,u_2)\,,
\end{equation}

\noindent
where

\begin{equation}
\delta^{14}(z_1-z_2) = \delta^6(x_1-x_2) \delta^8(\theta_1-\theta_2).
\end{equation}

It is convenient to multiply the identity (\ref{Ward_Identity}) with the analytic superfields $\lambda_1$, $q^+_2$, and $\widetilde q^+_3$, and integrate
the expression obtained over both analytic arguments,

\begin{eqnarray}\label{Ward_Identity_Integrated}
&& \int d\mu\, \widetilde q^+_3 D^{++}\lambda_1 q^+_2\, \frac{\delta^3\Delta\Gamma}{\delta V^{++}_1 \delta q^+_2 \delta\widetilde q^+_3} =  i \int d \zeta^{(-4)}_1 du_1\, d\zeta^{(-4)}_3\, du_3\,  \widetilde q^+_3 \lambda_1 q^+_1\, \frac{\delta^2\Delta\Gamma}{\delta q^+_1 \delta\widetilde q^+_3}\qquad\nonumber\\
&& - i \int d \zeta^{(-4)}_1 du_1\, d\zeta^{(-4)}_2\, du_2\, \widetilde q^+_1 \lambda_1 q^+_2\, \frac{\delta^2\Delta\Gamma}{\delta q^+_2 \delta\widetilde q^+_1},\qquad
\end{eqnarray}

\noindent
where

\begin{equation}
\int d\mu =  \int d\zeta^{(-4)}_1\, du_1\, d\zeta^{(-4)}_2\, du_2\, d\zeta^{(-4)}_3\, du_3.
\end{equation}

\noindent
This form of the Ward identity is most convenient, when checking it for one or another particular class of diagrams.

\subsection{The Feynman rules}
\hspace*{\parindent}

For the explicit calculation of quantum correction it is necessary to formulate the relevant Feynman rules. This can be accomplished quite similarly to the $4D$, ${\cal N}=2$
case  considered in detail in Refs. \cite{Galperin:1985bj,Galperin:1985va}. To find the propagator of the gauge superfield in the $\xi$-gauge, we consider the sum
of the gauge superfield action and the gauge-fixing term

\begin{eqnarray}\label{Gauge_Part_Of_Total_Action}
&& S_{\mbox{\scriptsize gauge}} + S_{\mbox{\scriptsize gf}} = \frac{1}{4f_0^2}\Big(1-\frac{1}{\xi_0}\Big) \int d^{14}z\, du_1 du_2 \frac{1}{(u_1^+ u_2^+)^2} V^{++}(z,u_1) V^{++}(z,u_2) \nonumber\\
&& + \frac{1}{4f_0^2\xi_0} \int d\zeta^{(-4)}\, du\, V^{++}(z,u) \partial^2 V^{++}(z,u), \qquad \label{Quadr}
\end{eqnarray}

\noindent
where we made use of the identity

\begin{equation}\label{U_Identity}
D_1^{++} \frac{1}{(u_1^+ u_2^+)^3} = \frac{1}{2} (D_1^{--})^2 \delta^{(3,-3)}(u_1,u_2)
\end{equation}

\noindent
and took into account that, when  acting on the analytic superfields,

\begin{equation}
\frac{1}{2} (D^+)^4 (D^{--})^2 \Rightarrow \partial^2.
\end{equation}

\noindent
Following Ref. \cite{Buchbinder:2017ozh}, we consider the free theory and solve the equation of motion for the superfield $V^{++}$ in the presence of the source term,

\begin{equation}
\frac{1}{2\xi_0 f_0^2} \partial^2 V^{++}(z,u_1) + \frac{1}{2f_0^2}\Big(1-\frac{1}{\xi_0}\Big) \int du_2 \frac{1}{(u_1^+ u_2^+)^2} (D_1^+)^4 V^{++}(z,u_2) + J^{(+2)}(z,u_1) = 0.
\end{equation}

\noindent
The solution can be presented as

\begin{equation}
V^{++}(z,u_1) = -\frac{2\xi_0 f_0^2}{\partial^2} J^{(+2)}(z,u_1) + \frac{2f_0^2(\xi_0-1)}{\partial^4} \int du_2 \frac{1}{(u_1^+ u_2^+)^2} (D_1^+)^4 J^{(+2)}(z,u_2),
\end{equation}

\noindent
whence one extracts the $\xi$-gauge form of the propagator of the gauge superfield

\begin{eqnarray}\label{Gauge_Propagator}
&& G_V^{(2,2)}(z_1,u_1;z_2,u_2) = - 2 f_0^2 \Big(\frac{\xi_0}{\partial^2} (D_1^+)^4 \delta^{(2,-2)}(u_2,u_1)\nonumber\\
&&\qquad\qquad\qquad\qquad\qquad - \frac{\xi_0-1}{\partial^4} (D_1^+)^4 (D_2^+)^4 \frac{1}{(u_1^+ u_2^+)^2}\Big) \delta^{14}(z_1-z_2).\qquad
\end{eqnarray}

\noindent
The second term vanishes in the Feynman gauge $\xi_0=1$. Such a choice considerably simplifies calculation of quantum corrections. However, the purpose of the present  paper  is to investigate the $\xi_0$-dependence of various Green functions for the generic choice of $\xi_0$.

\begin{figure}[h]
\begin{picture}(0,2.7)
\put(5,1.2){\includegraphics[scale=0.2]{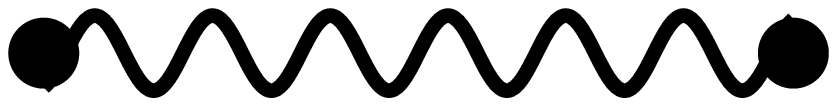}}
\put(4.4,0.1){(gauge multiplet)}
\put(9,1.2){\includegraphics[scale=0.2]{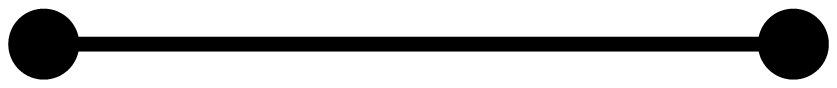}}
\put(8.5,0.1){(hypermultiplet)}
\end{picture}
\caption{The propagators of the gauge superfield $V^{++}$ and of the hypermultiplets.}\label{Figure_Propagators}
\end{figure}

In left part of Fig. \ref{Figure_Propagators}, the propagator (\ref{Gauge_Propagator}) is depicted by the wavy line ending on the points 1 and 2.

For completeness, we also present the expression for the hypermultiplet  propagator,

\begin{equation}\label{Hypermultiplet_Propagator}
G_q^{(1,1)}(z_1,u_1;z_2,u_2) = (D_1^+)^4 (D_2^+)^4 \frac{1}{\partial^2} \delta^{14}(z_1-z_2) \frac{1}{(u_1^+ u_2^+)^3},
\end{equation}

\noindent
which is denoted by the solid line on the right.

\bigskip

\begin{figure}[h]
\begin{picture}(0,2)
\put(7,0){\includegraphics[scale=0.15]{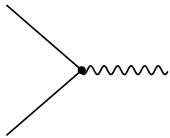}}
\end{picture}
\caption{The only vertex comes from the interaction of the hypermultiplet with the gauge superfield.}
\label{Figure_Vertex}
\end{figure}

The only vertex of the theory (\ref{Electodynamics_Action}) is presented in Fig. \ref{Figure_Vertex} and stands for the interaction
of the hypermultiplet with the gauge superfield

\begin{equation}\label{Vertex}
S_I = - i \int d\zeta^{(-4)}\, du\, \widetilde q^{+} V^{++} q^+.
\end{equation}

The superficial degree of divergence in the theory under consideration has been calculated in Ref. \cite{Buchbinder:2016gmc}:

\begin{equation}\label{Divergence_Degree}
\omega = 2L - N_q - \frac{1}{2} N_D.
\end{equation}

\noindent
Here $L$ is a number of loops, $N_q$ is a number of external hypermultiplet legs, and $N_D$ is a number of spinor supersymmetric covariant derivatives
acting on external legs. This formula implies that in the one-loop approximation only diagrams without external hypermultiplet legs or with two such legs can be divergent.

\section{Gauge dependence of the one-loop divergences}
\label{Section_Divergences}

\subsection{Two-point function of the gauge superfield}
\hspace*{\parindent}

\begin{figure}[h]
\begin{picture}(0,2)
\put(6.5,0){\includegraphics[scale=0.4]{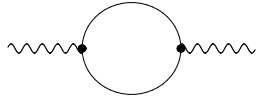}}
\end{picture}
\caption{The diagram  representing the one-loop two-point Green function in the abelian case.}
\label{Figure_Gauge_Diagram}
\end{figure}

In the one-loop approximation the two-point function of the gauge superfield $V^{++}$ is divergent.
In the abelian case this divergence comes only from the diagram pictured in Fig. \ref{Figure_Gauge_Diagram}.
However, this diagram does not contain propagators of the gauge superfield and is therefore gauge-independent.

Thus, in the one-loop approximation this Green function in the $\xi$-gauge is the same as in the Feynman gauge.
It is given by the expression \cite{Buchbinder:2016gmc}

\begin{equation}\label{Gauge_Contribution}
\int \frac{d^6p}{(2\pi)^6} \int d^8\theta\, du_1\, du_2\, V^{++}(p,\theta,u_1) V^{++}(-p,\theta,u_2) \frac{1}{(u_1^+ u_2^+)^2} \Big[\frac{1}{4f_0^2} - \frac{i}{2} \int \frac{d^6k}{(2\pi)^6} \frac{1}{k^2 (k+p)^2}\Big].\qquad
\end{equation}

The corresponding divergent part of the effective action is gauge-independent and in the dimensional  reduction scheme\footnote{Here we use
the regularization by dimensional reduction \cite{Siegel:1979wq}. However, for calculating power divergences
one should use another regularization, e.g., some modifications of the higher covariant derivative regularization \cite{Slavnov:1971aw,Slavnov:1972sq}.
At least for $4D$, ${\cal N}=2$ supersymmetric theories such a regularization can be formulated in the harmonic superspace \cite{Buchbinder:2015eva}.}
can be written as

\begin{equation}\label{Gauge_Divergence}
-\frac{1}{6\varepsilon (4\pi)^3}\int d\zeta^{(-4)}\, du\, (F^{++})^2,
\end{equation}

\noindent
where $\varepsilon = 6-D$.

\subsection{Two-point hypermultiplet Green function}
\hspace*{\parindent}

\begin{figure}[h]
\begin{picture}(0,2)
\put(6.5,0){\includegraphics[scale=0.4]{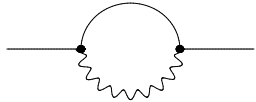}}
\end{picture}
\caption{The two-point Green function of the hypermultiplet in the one-loop approximation}\label{Figure_Hypermultiplet}
\end{figure}

In the one-loop approximation the two-point Green function of the hypermultiplet is contributed to by the single logarithmically divergent diagram
presented in Fig. \ref{Figure_Hypermultiplet}.

In the Feynman gauge this superdiagram vanishes. However, it includes the propagator of the gauge superfield, for which reason we can expect that the result
for it is in fact gauge-dependent. Using the Feynman rules defined above,  the expression for this diagram in the generic $\xi$-gauge can be written as

\begin{eqnarray}\label{Hypermultiplet_Diagram}
&& -2if_0^2 \int d\zeta^{(-4)}_1\, du_1\, d\zeta^{(-4)}_2\, du_2\, \widetilde q^+(z_1,u_1) q^+(z_2,u_2) \frac{1}{(u_1^+ u_2^+)^3} \frac{(D_1^+)^4 (D_2^+)^4}{\partial^2} \delta^{14}(z_1-z_2)\qquad\nonumber\\
&& \times \Big(\frac{\xi_0}{\partial^2} (D_1^+)^4 \delta^{(2,-2)}(u_2,u_1) - \frac{\xi_0-1}{\partial^4} (D_1^+)^4 (D_2^+)^4 \frac{1}{(u_1^+ u_2^+)^2}\Big) \delta^{14}(z_1-z_2).
\end{eqnarray}

\noindent
The derivatives $(D_1^+)^4 (D_2^+)^4$ in the hypermultiplet propagator can be used to convert the integrations over $d\zeta^{(-4)}$ into those over $d^{14}z$,

\begin{eqnarray}\label{Hypermultiplet_Green}
&& -2if_0^2 \int d^{14}z_1\, du_1\, d^{14}z_2\, du_2\, \widetilde q^+(z_1,u_1) q^+(z_2,u_2) \frac{1}{(u_1^+ u_2^+)^3}\,
\frac{1}{\partial^2} \delta^{14}(z_1-z_2)\qquad\nonumber\\
&& \times \Big(\frac{\xi_0}{\partial^2} (D_1^+)^4 \delta^{(2,-2)}(u_2,u_1) - \frac{\xi_0-1}{\partial^4} (D_1^+)^4 (D_2^+)^4 \frac{1}{(u_1^+ u_2^+)^2}\Big)  \delta^{14}(z_1-z_2).\label{Exp22}
\end{eqnarray}

\noindent
Taking into account the identities

\begin{eqnarray}\label{First_Identity}
&& \delta^{8}(\theta_1-\theta_2)\, (D^{+}_1)^4 \delta^{8}(\theta_1-\theta_2) =0,\vphantom{\Big(}\\
\label{Second_Identity}
&& \delta^{8}(\theta_1-\theta_2)\, (D^{+}_1)^4 (D^{+}_2)^4 \delta^{8}(\theta_1-\theta_2) = (u_1^+ u_2^+)^4\, \delta^{8}(\theta_1-\theta_2)\vphantom{\Big(}\,,\qquad
\end{eqnarray}

\noindent
we find that the first term in this expression vanishes, reducing (\ref{Hypermultiplet_Green}) to the form

\begin{equation}
2if_0^2 \int d^{6}x_1\, d^{6}x_2\,d^8\theta\, du_1\, du_2\, \widetilde q^+(x_1,\theta, u_1) q^+(x_2,\theta,u_2)
\frac{(\xi_0-1)}{(u_1^+ u_2^+)}\, \frac{1}{\partial^2} \delta^{6}(x_1-x_2)\,  \frac{1}{\partial^4} \delta^{6}(x_1-x_2).
\end{equation}

\noindent
This expression can be rewritten in the momentum representation as

\begin{equation}\label{Hypermultiplet_Green_Function}
- 2if_0^2 \int \frac{d^6p}{(2\pi)^6}\, \frac{d^6k}{(2\pi)^6} \frac{1}{k^4 (k+p)^2} \int d^8\theta\, du_1\, du_2\, \frac{(\xi_0-1)}{(u_1^+ u_2^+)} \widetilde q^+(p,\theta, u_1) q^+(-p,\theta\,u_2).
\end{equation}

\noindent
We observe that this expression is logarithmically divergent and does not vanish, unless the Feynman gauge is chosen.
If the theory is regularized by dimensional reduction,  the corresponding contribution to the divergent part takes the form

\begin{equation}\label{Two_Point_Function_Divergence}
-\frac{2f_0^2}{\varepsilon (4\pi)^3} \int d^{14}z\, du_1\, du_2\, \frac{(\xi_0-1)}{(u_1^+ u_2^+)} \widetilde q^+(z, u_1) q^+(z, u_2).
\end{equation}

\subsection{Three-point gauge-hypermultiplet Green function}
\hspace*{\parindent}

\begin{figure}[h]
\begin{picture}(0,2.5)
\put(6.5,0.){\includegraphics[scale=0.16]{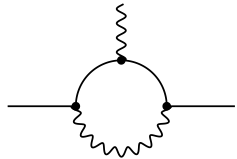}}
\end{picture}
\caption{The diagram representing the three-point gauge-hypermultiplet function in the one-loop approximation.}\label{Figure_One-Loop_Vertex}
\end{figure}

According to Eq. (\ref{Divergence_Degree}), all diagrams containing two external hypermultiplet legs are logarithmically divergent,
irrespective of the number of the external gauge legs. That is why in calculating the one-loop divergences it is necessary to take into account
such Green functions. The simplest of them is the three-point gauge superfield - hypermultiplet. In the one-loop approximation, it is contributed to
by the single supergraph  depicted in Fig. \ref{Figure_One-Loop_Vertex}.

Calculating this diagram by  Feynman rules in the general $\xi$-gauge, we obtain

\begin{eqnarray}
&& - 2f_0^2 \int d\zeta^{(-4)}_1\,du_1\,d\zeta^{(-4)}_2\,du_2\,d\zeta^{(-4)}_3\,du_3\,\widetilde q^+(z_1,u_1)
q^+(z_3,u_3) V^{++}(z_2,u_2) \Big(\frac{\xi_0}{\partial^2} (D_1^+)^4 \qquad\nonumber\\
&&\times \delta^{(2,-2)}(u_3,u_1) - \frac{(\xi_0-1)}{\partial^4} (D_1^+)^4 (D_3^+)^4 \frac{1}{(u_1^+ u_3^+)^2}\Big) \delta^{14}(z_1-z_3)\, \frac{1}{(u_1^+ u_2^+)^3}\frac{(D_1^+)^4 (D_2^+)^4}{\partial^2}\qquad\nonumber\\
&&\times \delta^{14}(z_1-z_2)\, \frac{1}{(u_2^+ u_3^+)^3}\frac{(D_2^+)^4 (D_3^+)^4}{\partial^2} \delta^{14}(z_2-z_3).\label{Expr33}
\end{eqnarray}

\noindent
To work out this expression, we, first, convert the integrals over $d\zeta^{(-4)}$ in it into integrals over $d^{14}z$ using Eq. (\ref{Measure_Analytic}):

\begin{eqnarray}
&& - 2f_0^2 \int d^{14}z_1\,d^{14}z_2\,d^{14}z_3\,du_1\,du_2\,du_3\,\widetilde q^+(z_1,u_1) q^+(z_3,u_3) V^{++}(z_2,u_2)
\Big(\frac{\xi_0}{\partial^2} (D_1^+)^4  \qquad\nonumber\\
&& \times \delta^{(2,-2)}(u_3,u_1) - \frac{(\xi_0-1)}{\partial^4} (D_1^+)^4 (D_3^+)^4 \frac{1}{(u_1^+ u_3^+)^2}\Big)
\delta^{14}(z_1-z_3)\, \frac{1}{(u_1^+ u_2^+)^3 (u_2^+ u_3^+)^3}
\qquad \nonumber\\
&& \times \frac{(D_2^+)^4}{\partial^2} \delta^{14}(z_1-z_2)\, \frac{1}{\partial^2} \delta^{14}(z_2-z_3).\vphantom{\frac{1}{2}}
\end{eqnarray}

\noindent
Next, we integrate by parts with respect to  $(D^+_2)^4$ (assuming that $D_2^+$ acts on $z_1$), taking into account that

\begin{equation}
\delta^8(\theta_1-\theta_2)\prod\limits_{n=1}^N D^+_{i_n a_n} \delta^8(\theta_1-\theta_2) = 0 \qquad \mbox{for arbitrary odd $N$}.
\end{equation}

\noindent
In the term containing the harmonic $\delta$-function we further integrate over $du_3$. Integrating also over $\theta_2$,  we finally obtain for (\ref{Expr33}):

\begin{eqnarray}
&&  2f_0^2 \int d^{6}x_1\,d^{6}x_2\,d^{6}x_3\,d^8\theta_1\,d^8\theta_3\,\delta^8(\theta_1-\theta_3)\, \Bigg\{
\int du_1\, du_2\, \widetilde q^+(x_1,\theta_1,u_1) q^+(x_3,\theta_3,u_1) \nonumber\\
&& \times V^{++}(x_2,\theta_1, u_2) \frac{\xi_0}{(u_1^+ u_2^+)^6}\,   \frac{(D_1^+)^4 (D_2^+)^4}{\partial^2} \delta^{14}(z_1-z_3)\,\frac{1}{\partial^2} \delta^{6}(x_1-x_2)\,
\,\frac{1}{\partial^2} \delta^{6}(x_2-x_3)
\nonumber\\
&& + \int du_1\,du_2\,du_3\,  V^{++}(x_2,\theta_1, u_2) q^+(x_3,\theta_3,u_3) \frac{(\xi_0-1)}{(u_1^+ u_3^+)^2 (u_1^+ u_2^+)^3 (u_2^+ u_3^+)^3}\frac{1}{\partial^2} \delta^{6}(x_1-x_2) \qquad\nonumber\\
&&\times \frac{1}{\partial^2} \delta^{6}(x_2-x_3)\, \Bigg[(D_2^+)^4 \widetilde q^+(x_1,\theta_1,u_1)\, \frac{(D_1^+)^4 (D_3^+)^4}{\partial^4} \delta^{14}(z_1-z_3)
 + \widetilde q^+(x_1,\theta_1,u_1)\nonumber\\
&&\times \frac{(D_2^+)^4 (D_1^+)^4 (D_3^+)^4}{\partial^4} \delta^{14}(z_1-z_3)  - \frac{1}{4}\varepsilon^{abcd} D_{2a}^+ D_{2b}^+\, \widetilde q^+(x_1,\theta_1,u_1)\, \frac{D_{2c}^+ D_{2d}^+ (D_1^+)^4 (D_3^+)^4}{\partial^4}  \nonumber\\
&&\times \delta^{14}(z_1-z_3)\Bigg] \Bigg\}.
\end{eqnarray}

\noindent
As the further step,  we use the identities (\ref{First_Identity}), (\ref{Second_Identity}) together with

\begin{eqnarray}\label{Third_Identity}
&& \delta^{8}(\theta_1-\theta_2)\, D^{+}_{2a} D^{+}_{2b} (D^{+}_1)^4 (D^{+}_3)^4\delta^{8}(\theta_1-\theta_2)\vphantom{\Big(}\nonumber\\
&&\qquad\qquad\qquad\qquad = -i(\gamma^M)_{ab} (u_2^+ u_1^+)\, (u_2^+ u_3^+)\, (u_1^+ u_3^+)^3 \,\delta^{8}(\theta_1-\theta_2) \partial_M;\vphantom{\Big(}\qquad\\
\label{Fifth_Identity}
&& \delta^{8}(\theta_1-\theta_2)\, (D^{+}_2)^4 (D^{+}_1)^4 (D^{+}_3)^4\delta^{8}(\theta_1-\theta_2)\vphantom{\Big(}\nonumber\\
&&\qquad\qquad\qquad\qquad\qquad = (u_1^+ u_2^+)^2\, (u_1^+ u_3^+)^2\, (u_2^+ u_3^+)^2\,\delta^{8}(\theta_1-\theta_2) \partial^2 \vphantom{\Big(}\qquad
\end{eqnarray}

\noindent
in order to do the integrals over the Grassmann coordinate $\theta_2$. After renaming $\theta_1 \to \theta$, the expression for the diagram in question
in the momentum representation is written as

\begin{eqnarray}\label{Three-Point_Function}
&& 2f_0^2 \int \frac{d^{6}p}{(2\pi)^6}\,\frac{d^{6}q}{(2\pi)^6}\,\frac{d^{6}k}{(2\pi)^6}\,d^8\theta\,\Bigg\{-\int du_1\,du_2\,\widetilde q^+(q+p,\theta,u_1) V^{++}(-p,\theta,u_2) q^+(-q,\theta,u_1)
\nonumber\\
&& \times \frac{\xi_0}{k^2 (q+k)^2 (q+k+p)^2} \frac{1}{(u_1^+ u_2^+)^2}
+ \int du_1\,du_2\,du_3\,\Bigg[ (D^+_{2})^4\,\widetilde q^+(q+p,\theta,u_1)\,V^{++}(-p,\theta, u_2)\nonumber\\
&&\times  q^+(-q,\theta,u_3)\,
\frac{(\xi_0-1)}{k^4 (q+k)^2 (q+k+p)^2}\, \frac{(u_1^+ u_3^+)^2}{(u_1^+ u_2^+)^3 (u_2^+ u_3^+)^3}
- \widetilde q^+(q+p,\theta,u_1)\, V^{++}(-p,\theta, u_2)\nonumber\\
&& \times q^+(-q,\theta,u_3)
\frac{(\xi_0-1)}{k^2 (q+k)^2 (q+k+p)^2} \frac{1}{(u_1^+ u_2^+) (u_2^+ u_3^+)}
- D^+_{2a} D^+_{2b}\,\widetilde q^+(q+p,\theta,u_1)\nonumber\\
&&\times V^{++}(-p,\theta, u_2)\, q^+(-q,\theta,u_3)\,
\frac{(\xi_0-1)(\widetilde\gamma^M)^{ab} k_M}{2 k^4 (q+k)^2 (q+k+p)^2}\, \frac{(u_1^+ u_3^+)}{(u_1^+ u_2^+)^2 (u_2^+ u_3^+)^2}
\Bigg]\Bigg\},
\end{eqnarray}

\noindent
where $(\widetilde\gamma^M)^{ab} = \varepsilon^{abcd} (\gamma^M)_{cd}/2$. The divergent part of this expression can now be found after the Wick rotation.
There remains only one divergent integral

\begin{equation}
\int \frac{d^6k}{(2\pi)^6}\frac{1}{k^2 (k+q)^2 (k+q+p)^2}\,,
\end{equation}

\noindent
which, after regularizing it by dimensional reduction, is reduced to

\begin{equation}
-i\int \frac{d^DK}{(2\pi)^6}\frac{1}{K^2 (K+Q)^2 (K+Q+P)^2} = -\frac{i}{\varepsilon (4\pi)^3} + \mbox{finite terms},
\end{equation}

\noindent
where the capital letters denote Euclidean momentums. Thus, the divergent part of the diagram in Fig. 5
can be presented as

\begin{equation}\label{Three_Point_Function_Divergence}
\frac{2if_0^2}{\varepsilon (4\pi)^3} \int d^{14}z \Bigg\{\int du_1\,du_2\,\widetilde q^+_1 V^{++}_2 q^+_1
\frac{\xi_0}{(u_1^+ u_2^+)^2}
+ \int du_1\,du_2\,du_3\, \widetilde q^+_1\, V^{++}_2 q^+_3 \frac{(\xi_0-1)}{(u_1^+ u_2^+) (u_2^+ u_3^+)}
\Bigg\},
\end{equation}

\noindent
where the subscripts on the superfields refer to the relevant harmonic arguments.

\section{Verification of the Ward identities}
\hspace*{\parindent}\label{Section_Ward}

To be convinced of the correctness of the results obtained in the previous sections, let us check that the two- and three-point Green functions derived above satisfy the Ward identities.

First, for completeness, we verify the Ward identity (\ref{Transversality}). The two-point Green function of the gauge superfield is obtained
by differentiating Eq. (\ref{Gauge_Contribution}) with respect to   $V^{++}$, using Eq. (\ref{Analytic_Derivative}). This gives

\begin{equation}
\frac{\delta^2\Delta\Gamma}{\delta V^{++}_1 \delta V^{++}_2} = G_V(i\partial_M) \frac{1}{(u_1^+ u_2^+)^2} (D_1^{+})^4 (D_2^+)^4 \delta^{14}(z_1-z_2),
\end{equation}

\noindent
where

\begin{equation}
G_V(p_M) = \frac{1}{2f_0^2} - i \int \frac{d^6k}{(2\pi)^6}\frac{1}{k^2 (k+p)^2} + \ldots
\end{equation}

\noindent
Therefore,

\begin{eqnarray}
&&\hspace*{-6mm} D_1^{++}\frac{\delta^2\Delta\Gamma}{\delta V^{++}_1 \delta V^{++}_2} = G_V(i\partial_M) D_1^{--} \delta^{(2,-2)}(u_1,u_2)\cdot (D_1^{+})^4 (D_2^+)^4 \delta^{14}(z_1-z_2) = G_V(i\partial_M)\nonumber\\
&&\hspace*{-6mm} \times \Big[D_1^{--} \Big(\delta^{(2,-2)}(u_1,u_2) (D_1^{+})^4 (D_2^+)^4 \Big) - \delta^{(2,-2)}(u_1,u_2) \Big(D_1^{--} (D_1^{+})^4\Big) (D_2^+)^4\Big]\delta^{14}(z_1-z_2) = 0.\nonumber\\
\end{eqnarray}

\noindent
Thus, we have verified that the Ward identity (\ref{Transversality}) is indeed satisfied.

The two-point Green function of the hypermultiplet is obtained by differentiating Eq. (\ref{Hypermultiplet_Green_Function}) with respect to $q^+$ and $\widetilde q^+$.
These derivatives are calculated  with the help of Eq. (\ref{Analytic_Derivative}). We obtain

\begin{equation}
\frac{\delta^2\Gamma}{\delta q_2^+\,\delta \widetilde q_1^+} = G_q(i\partial_M) \frac{1}{(u_1^+ u_2^+)} (D^+_1)^4 (D^+_2)^4 \delta^{14}(z_1-z_2),
\end{equation}

\noindent
where

\begin{equation}
G_q(p_M) = - 2if_0^2 \int \frac{d^6k}{(2\pi)^6} \frac{(\xi_0-1)}{k^4 (k+p)^2} +\ldots
\end{equation}

The three-point gauge superfield - hypermultiplet Green function can be constructed quite similarly, starting from Eq. (\ref{Three-Point_Function}),
but we prefer not to present the expression for it explicitly. Instead, we will check for it the Ward identity in the form (\ref{Ward_Identity_Integrated}).
From Eq. (\ref{Three-Point_Function}) we obtain

\begin{eqnarray}\label{Three-Point_With_Lambda}
&& \int d\zeta^{(-4)}_1\,du_1\, d\zeta^{(-4)}_2\,du_2\, d\zeta^{(-4)}_3\,du_3\, \widetilde q^+_3 D^{++}\lambda_1 q^+_2\, \frac{\delta^3\Delta\Gamma}{\delta V^{++}_1 \delta q^+_2 \delta\widetilde q^+_3} \nonumber\\
&& = 2f_0^2 \int \frac{d^{6}p}{(2\pi)^6}\,\frac{d^{6}q}{(2\pi)^6}\,\frac{d^{6}k}{(2\pi)^6}\,d^8\theta\,\Bigg\{-\int du_1\,du_2\,\widetilde q^+(q+p,\theta,u_2) D^{++}_1\lambda(-p,\theta,u_1) q^+(-q,\theta,u_2)
\nonumber\\
&& \times \frac{\xi_0}{k^2 (q+k)^2 (q+k+p)^2} \frac{1}{(u_1^+ u_2^+)^2}
+ \int du_1\,du_2\,du_3\,\Bigg[ (D^+_{1})^4\,\widetilde q^+(q+p,\theta,u_3)\,D^{++}_1\lambda(-p,\theta, u_1)\nonumber\\
&&\times  q^+(-q,\theta,u_2)\,
\frac{(\xi_0-1)}{k^4 (q+k)^2 (q+k+p)^2}\, \frac{(u_3^+ u_2^+)^2}{(u_3^+ u_1^+)^3 (u_1^+ u_2^+)^3}
- \widetilde q^+(q+p,\theta,u_3)\, D^{++}_1\lambda(-p,\theta, u_1)\nonumber\\
&& \times q^+(-q,\theta,u_2)
\frac{(\xi_0-1)}{k^2 (q+k)^2 (q+k+p)^2} \frac{1}{(u_3^+ u_1^+) (u_1^+ u_2^+)}
- D^+_{1a} D^+_{1b}\,\widetilde q^+(q+p,\theta,u_3)\nonumber\\
&&\times D^{++}_1\lambda(-p,\theta, u_1)\, q^+(-q,\theta,u_2)\,
\frac{(\xi_0-1)(\widetilde\gamma^M)^{ab} k_M}{2 k^4 (q+k)^2 (q+k+p)^2}\, \frac{(u_3^+ u_2^+)}{(u_3^+ u_1^+)^2 (u_1^+ u_2^+)^2}
\Bigg]\Bigg\}.
\end{eqnarray}

\noindent
Next, we integrate by parts with respect to the harmonic derivatives $D^{++}_1$, taking into account the identity

\begin{equation}\label{Harmonic_Identity}
D^{++}_1 \frac{1}{(u_1^+ u_2^+)^n} = \frac{1}{(n-1)!} (D_1^{--})^{n-1}\delta^{(n,-n)}(u_1,u_2) = \frac{(-1)^{n-1}}{(n-1)!} (D_2^{--})^{n-1}\delta^{(2-n,n-2)}(u_1,u_2) .
\end{equation}

\noindent
After some algebra (described in Appendix \ref{Appendix_Ward}), this gives

\begin{eqnarray}\label{Three-Point_Ward}
&&\hspace*{-6mm} \int d\mu\, \widetilde q^+_3 D^{++}\lambda_1 q^+_2\, \frac{\delta^3\Delta\Gamma}{\delta V^{++}_1 \delta q^+_2 \delta\widetilde q^+_3} = - 2f_0^2 \int \frac{d^6p}{(2\pi)^6}\, \frac{d^6q}{(2\pi)^6}\, \frac{d^6k}{(2\pi)^6} \frac{1}{k^4 (k+q+p)^2} \int d^8\theta\, du_1\, du_3\,\qquad\nonumber\\
&&\hspace*{-6mm} \times \frac{(\xi_0-1)}{(u_1^+ u_3^+)} \widetilde q^+(q+p,\theta, u_3) \lambda(-p,\theta,u_1) q^+(-q,\theta\,u_1) - 2f_0^2 \int \frac{d^6p}{(2\pi)^6}\, \frac{d^6q}{(2\pi)^6}\, \frac{d^6k}{(2\pi)^6} \frac{1}{k^4 (k+q)^2} \nonumber\\
&&\hspace*{-6mm} \times \int d^8\theta\, du_1\, du_2\, \frac{(\xi_0-1)}{(u_1^+ u_2^+)} \widetilde q^+(q+p,\theta, u_1) \lambda(-p,\theta,u_1) q^+(-q,\theta\,u_2).
\end{eqnarray}

\noindent
The right-hand side of this equation can be rewritten as

\begin{eqnarray}
&&  i \int d \zeta^{(-4)}_1 du_1\, d\zeta^{(-4)}_3\, du_3\,  \widetilde q^+_3 \lambda_1 q^+_1\, \frac{\delta^2\Gamma}{\delta q^+_1 \delta\widetilde q^+_3}
\nonumber\\
&& - i \int d \zeta^{(-4)}_1 du_1\, d\zeta^{(-4)}_2\, du_2\, \widetilde q^+_1 \lambda_1 q^+_2\, \frac{\delta^2\Gamma}{\delta q^+_2 \delta\widetilde q^+_1},\qquad
\end{eqnarray}

\noindent
thus demonstrating that the Green functions (\ref{Hypermultiplet_Green_Function}) and (\ref{Three-Point_Function}) satisfy the Ward identity (\ref{Ward_Identity_Integrated}), as
it should be. Obviously, they also satisfy the Ward identity in the original form (\ref{Ward_Identity}).
This completes checking the correctness of our calculation.

\section{The vanishing of the gauge dependence on shell}
\hspace*{\parindent}\label{Section_Shell}

According to the general theorem of Refs. \cite{DeWitt:1965jb,Boulware:1980av,Voronov:1981rd,Voronov:1982ph,Voronov:1982ur,Lavrov:1986hr}, the gauge-dependent terms
should disappear on shell. Let us verify that our results are in agreement with this statement.

It is convenient to represent the effective action in the form

\begin{equation}
\Gamma = \Gamma_{\xi_0=1} + \widetilde\Gamma,
\end{equation}

\noindent
where

\begin{eqnarray}
&& \Gamma_{\xi_0=1} = S + S_{\mbox{\scriptsize gf}} -\frac{i}{2} \int \frac{d^6p}{(2\pi)^6} \int d^8\theta\, du_1\, du_2\, V^{++}(p,\theta,u_1) V^{++}(-p,\theta,u_2) \frac{1}{(u_1^+ u_2^+)^2}  \nonumber\\
&&\times  \int \frac{d^6k}{(2\pi)^6}\frac{1}{k^2 (k+p)^2}
- \int \frac{d^{6}p}{(2\pi)^6}\,\frac{d^{6}q}{(2\pi)^6}\,d^8\theta\, du_1\,du_2\,\widetilde q^+(q+p,\theta,u_1) V^{++}(-p,\theta,u_2) \qquad
\nonumber\\
&& \times q^+(-q,\theta,u_1) \frac{1}{(u_1^+ u_2^+)^2}\int\frac{d^{6}k}{(2\pi)^6}\, \frac{2f_0^2}{k^2 (q+k)^2 (q+k+p)^2}  + \ldots
\end{eqnarray}

\noindent
is the effective action in the Feynman gauge and

\begin{eqnarray}\label{Gamma_Gauge_Dependent}
&& \widetilde\Gamma =
- 2if_0^2 \int \frac{d^6p}{(2\pi)^6}\, \frac{d^6k}{(2\pi)^6} \frac{1}{k^4 (k+p)^2} \int d^8\theta\, du_1\, du_2\, \frac{(\xi_0-1)}{(u_1^+ u_2^+)} \widetilde q^+(p,\theta, u_1)\, q^+(-p,\theta\,u_2)\nonumber\\
&& + 2f_0^2 \int \frac{d^{6}p}{(2\pi)^6}\,\frac{d^{6}q}{(2\pi)^6}\,\frac{d^{6}k}{(2\pi)^6}\,d^8\theta\,\Bigg\{-\int du_1\,du_2\,\widetilde q^+(q+p,\theta,u_1) V^{++}(-p,\theta,u_2) q^+(-q,\theta,u_1)
\nonumber\\
&& \times \frac{(\xi_0-1)}{k^2 (q+k)^2 (q+k+p)^2} \frac{1}{(u_1^+ u_2^+)^2}
+ \int du_1\,du_2\,du_3\,\Bigg[ (D^+_{2})^4\,\widetilde q^+(q+p,\theta,u_1)\,V^{++}(-p,\theta, u_2)\nonumber\\
&&\times  q^+(-q,\theta,u_3)\,
\frac{(\xi_0-1)}{k^4 (q+k)^2 (q+k+p)^2}\, \frac{(u_1^+ u_3^+)^2}{(u_1^+ u_2^+)^3 (u_2^+ u_3^+)^3}
- \widetilde q^+(q+p,\theta,u_1)\, V^{++}(-p,\theta, u_2)\nonumber\\
&& \times q^+(-q,\theta,u_3)
\frac{(\xi_0-1)}{k^2 (q+k)^2 (q+k+p)^2} \frac{1}{(u_1^+ u_2^+) (u_2^+ u_3^+)}
- D^+_{2a} D^+_{2b}\,\widetilde q^+(q+p,\theta,u_1)\nonumber\\
&&\times V^{++}(-p,\theta, u_2)\, q^+(-q,\theta,u_3)\,
\frac{(\xi_0-1)(\widetilde\gamma^M)^{ab} k_M}{2k^4 (q+k)^2 (q+k+p)^2}\, \frac{(u_1^+ u_3^+)}{(u_1^+ u_2^+)^2 (u_2^+ u_3^+)^2}
\Bigg]\Bigg\}
+ \ldots
\end{eqnarray}

\noindent
stands for the gauge-dependent remainder of the effective action.

The purpose of this section is to demonstrate, by an explicit calculation, that in the  approximation considered, $\widetilde\Gamma$
indeed vanishes on shell. To this end, we use the equations of motion for the hypermultiplets following from the action (\ref{Electodynamics_Action}),

\begin{equation}
0 = \nabla^{++} q^+ = D^{++} q^+ + iV^{++} q^+;\qquad 0 = \nabla^{++} \widetilde q^+ = D^{++} \widetilde q^+ - iV^{++} \widetilde q^+.
\end{equation}

\noindent
In Appendix \ref{Appendix_Shell} (after some lengthy calculations) we demonstrate that, with these equations taken into account, the gauge-dependent part
of the one-loop effective action can be  cast in the form

\begin{eqnarray}\label{On_Shell_Result}
&& \widetilde\Gamma = 2f_0^2 \int \frac{d^{6}p}{(2\pi)^6}\,\frac{d^{6}q}{(2\pi)^6}\,d^4\theta^+\,du\, \widetilde q^+(q+p,\theta,u) V^{++}(-p,\theta,u) q^+(-q,\theta,u)
\nonumber\\
&& \times  \Big((q+p)^2+q^2\Big) \int \frac{d^{6}k}{(2\pi)^6}\, \frac{(\xi_0-1)}{k^2 (q+k)^2 (q+k+p)^2} + O\left((V^{++})^2\right).\qquad
\end{eqnarray}

\noindent
On shell, where $q^2=0$ and $(q+p)^2=0$,\footnote{These equations can be derived directly from the hypermultiplet free equation of motion, see Ref. \cite{Bossard:2015dva}
for details.} this expression vanishes. Thereby we have proved  that the gauge dependence is vanishing on shell.

Note that, while deriving this result, we ignored all terms proportional to $(V^{++})^k$ for $k\ge 2$, because in this paper
we limit our attention only to the diagrams without external gauge superfield legs  at all, and to those having a single gauge superfield leg.
In this approximation, terms of higher orders in $V^{++}$ are irrelevant.

\section{The total divergent part of the one-loop effective action}
\hspace{\parindent}\label{Section_Total_Divergences}

So far we investigated gauge dependence of the two- and three-point Green functions only. In particular,
we demonstrated that the corresponding one-loop divergences are gauge-dependent. However, according to Eq. (\ref{Divergence_Degree}),
the Green functions with an arbitrary number of external gauge legs (and two external hypermultiplet legs) are also divergent.
Nevertheless, the total divergent part of the one-loop effective action can be found using the reasoning based on the gauge invariance.
Actually, the one-loop divergences corresponding to the two- and three-point Green functions (see Eqs. (\ref{Gauge_Divergence}), (\ref{Two_Point_Function_Divergence}),
and (\ref{Three_Point_Function_Divergence})) have the form

\begin{eqnarray}\label{Lowest_Divergences}
&& \Gamma^{(1)}_\infty = -\frac{1}{6\varepsilon (4\pi)^3}\int d\zeta^{(-4)}\, du\, (F^{++})^2
-\frac{2f_0^2}{\varepsilon (4\pi)^3} \int d^{14}z\, du_1\, du_2\, \frac{(\xi_0-1)}{(u_1^+ u_2^+)} \widetilde q^+_1 q^+_2
+ \frac{2if_0^2}{\varepsilon (4\pi)^3}
\nonumber\\
&& \times \int d^{14}z \Bigg\{\int du_1\,du_2\,\widetilde q^+_1 V^{++}_2 q^+_1
\frac{\xi_0}{(u_1^+ u_2^+)^2}
+ \int du_1\,du_2\,du_3\, \widetilde q^+_1\, V^{++}_2 q^+_3 \frac{(\xi_0-1)}{(u_1^+ u_2^+) (u_2^+ u_3^+)}
\Bigg\}\qquad\nonumber\\
&& + O\Big(\widetilde q^+ (V^{++})^2 q^+\Big).\vphantom{\frac{1}{2}}
\end{eqnarray}

\noindent
The first term in this equation is gauge invariant. The expression corresponding to the first term in the curly brackets can also be  rewritten in the explicitly gauge invariant form,

\begin{eqnarray}
&& \frac{2if_0^2}{\varepsilon (4\pi)^3} \int d^{14}z\, du_1\,du_2\,\widetilde q^+_1 V^{++}_2 q^+_1
\frac{\xi_0}{(u_1^+ u_2^+)^2} = \xi_0\, \frac{2if_0^2}{\varepsilon (4\pi)^3} \int d^{14}z\, du\, \widetilde q^+ V^{--} q^+\nonumber\\
&& =  \xi_0\,\frac{2if_0^2}{\varepsilon (4\pi)^3} \int d\zeta^{(-4)}\, du\, \widetilde q^+ F^{++} q^+.
\end{eqnarray}

\noindent
According to Eq. (\ref{Q-_Expansion}), the remaining two  terms in Eq. (\ref{Lowest_Divergences}) are the lowest terms in the series expansion of the gauge invariant expression

\begin{equation}
- \frac{2 f_0^2 (\xi_0-1)}{\varepsilon (4\pi)^3} \int d^{14}z\, du\, \widetilde q^+\, q^-
\end{equation}

\noindent
in powers of $V^{++}$. Thus, the divergent part of the one-loop effective action can be written as

\begin{eqnarray}\label{One-Loop_Divergence}
&& \Gamma^{(1)}_\infty = -\frac{1}{6\varepsilon (4\pi)^3}\int d\zeta^{(-4)}\, du\, (F^{++})^2 + \frac{2if_0^2 \xi_0}{\varepsilon (4\pi)^3} \int d\zeta^{(-4)}\, du\, \widetilde q^+ F^{++} q^+\nonumber\\
&& - \frac{2 f_0^2 (\xi_0-1)}{\varepsilon (4\pi)^3} \int d^{14}z\, du\, \widetilde q^+\, q^-.
\end{eqnarray}

\noindent
Note that this expression does not include $O\Big(\widetilde q^+ (V^{++})^2 q^+\Big)$, because for obtaining
the gauge invariant expression such terms should contain $F^{++}$ in which the number $N_D=4$ of spinor derivatives acts on $V^{--}$. However,
according to Eq. (\ref{Divergence_Degree}) these terms are finite and do not contribute to the divergent part of the one-loop effective action.
Therefore, Eq. (\ref{One-Loop_Divergence}) provides the exact result for the divergent part of the effective action of the theory in question.

Note that on shell the gauge dependence of Eq. (\ref{One-Loop_Divergence}) vanishes. Actually, on shell, as the consequence of the equation of motion $\nabla^{++}q^+ = 0\,$, we have
the chain of relations

\begin{equation}
(\nabla^{++})^2 q^- = 0 \;\Rightarrow\;  (\nabla^{++})^2 \nabla^{--}q^- = 0 \; \Rightarrow\; \nabla^{++} \nabla^{--} q^- = 0 \; \Rightarrow\;  \nabla^{--} q^- = 0\,.
\end{equation}

Acting on the latter equation by $\nabla^{++}$ it is easy to find
\begin{equation}
q^- = \nabla^{--} q^+. \label{q-q+2}
\end{equation}

\noindent
In deriving these relations, we made use of the well known properties $D^{++}\omega^{-n} = 0 \rightarrow \omega^{-n} = 0,\; D^{--}\omega^{+ m} = 0 \rightarrow \omega^{+ m} = 0$
for $n\geq 1, m\geq 1$.

As a consequence of (\ref{q-q+2}), we obtain that on shell

\begin{eqnarray}
&& \int d^{14}z\, du\, \widetilde q^+\, q^- = \int d\zeta^{(-4)}\, du\, (D^+)^4\Big(\widetilde q^+\, \nabla^{--} q^+\Big)\nonumber\\
&&\qquad\qquad = \int d\zeta^{(-4)}\, du\, \widetilde q^+\, (D^+)^4\Big( (D^{--} +iV^{--}) q^+\Big) = i \int d\zeta^{(-4)}\, du\, \widetilde q^+\, F^{++} q^+.\qquad
\end{eqnarray}

\noindent
Thus, on shell, the one-loop divergence (\ref{One-Loop_Divergence}) takes the form

\begin{equation}
\Gamma^{(1)}_\infty = -\frac{1}{6\varepsilon (4\pi)^3}\int d\zeta^{(-4)}\, du\, (F^{++})^2 + \frac{2if_0^2}{\varepsilon (4\pi)^3} \int d\zeta^{(-4)}\, du\, \widetilde q^+ F^{++} q^+.
\end{equation}

\noindent
We see that this expression does not depend on the parameter $\xi$ and, hence, on the gauge choice.

\section{Summary}
\hspace*{\parindent}\label{Section_Summary}

In this paper, using the $6D\,,$ ${\cal N}=(1,0)$ harmonic superspace formalism, we studied the gauge dependence
of the one-loop effective action for ${\cal N}=(1,0)$ supersymmetric quantum electrodynamics. As compared to the case of the Feynman gauge,
in the general $\xi$-gauge some new divergences appear. In particular, we demonstrated that in the general case the hypermultiplet
Green function is divergent already in the one-loop approximation, as opposed to the case of the Feynman gauge, in which this divergence
vanishes. Moreover, we calculated the three-point gauge - hypermultiplet Green function in the general $\xi$-gauge.
To check the correctness of the calculation, we have verified the relevant Ward identity. Also it was checked that the gauge dependence vanishes on shell.
Taking into account the gauge invariance,  we also restored the divergent part of the one-loop effective action with terms of higher
orders in the gauge superfield $V^{++}$. It is given by Eq. (\ref{One-Loop_Divergence}) and contains a new term which is absent in the Feynman gauge. We demonstrated that the gauge dependence of this general expression also vanishes on shell.

It would be interesting to investigate the gauge dependence in the non-abelian case. In particular, from the results of this paper
we can expect that in the general $\xi$-gauge the $6D\,,$ ${\cal N}=(1,1)$ sypersymmetric Yang--Mills theory is not finite even
in the one-loop approximation, while the divergent terms are vanishing on shell.

\section*{Acknowledgements} \
\hspace*{\parindent}

This work was supported by the grant of Russian Science Foundation, project No. 16-12-10306.

\appendix

\section{Ward identity in harmonic superspace}
\hspace*{\parindent}\label{Appendix_Ward}

Let us show how to pass from  Eq. (\ref{Three-Point_With_Lambda}) to its equivalent  form (\ref{Three-Point_Ward}). After integrating by parts with respect to the derivatives $D^{++}_1$
 and using the identity (\ref{Harmonic_Identity}), we obtain

\begin{eqnarray}
&& \int d\mu\, \widetilde q^+_3 D^{++}\lambda_1 q^+_2\, \frac{\delta^3\Delta\Gamma}{\delta V^{++}_1 \delta q^+_2 \delta\widetilde q^+_3}\nonumber\\
&& = 2f_0^2 \int \frac{d^{6}p}{(2\pi)^6}\,\frac{d^{6}q}{(2\pi)^6}\,\frac{d^{6}k}{(2\pi)^6}\,d^8\theta\,\Bigg\{-\int du_1\,du_2\,\widetilde q^+(q+p,\theta,u_2) \lambda(-p,\theta,u_1) q^+(-q,\theta,u_2)
\nonumber\\
&& \times \frac{\xi_0}{k^2 (q+k)^2 (q+k+p)^2} D_2^{--} \delta^{(0,0)}(u_1,u_2)
+ \int du_1\,du_2\,du_3\,\Bigg[ (D^+_{1})^4\,\widetilde q^+(q+p,\theta,u_3)\nonumber\\
&&\times  \lambda(-p,\theta, u_1) q^+(-q,\theta,u_2)\,
\frac{(\xi_0-1)(u_3^+ u_2^+)^2}{k^4 (q+k)^2 (q+k+p)^2}\, \Bigg( \frac{1}{2 (u_1^+ u_2^+)^3} (D^{--}_3)^2 \delta^{(-1,1)}(u_1,u_3)\nonumber\\
&& + \frac{1}{2 (u_1^+ u_3^+)^3} (D^{--}_2)^2 \delta^{(-1,1)}(u_1,u_2) \Bigg)
- \widetilde q^+(q+p,\theta,u_3)\, \lambda(-p,\theta, u_1) q^+(-q,\theta,u_2) \nonumber\\
&&\times \frac{(\xi_0-1)}{k^2 (q+k)^2 (q+k+p)^2}\Bigg(\frac{1}{(u_1^+ u_3^+)} \delta^{(1,-1)}(u_1,u_2) + \delta^{(1,-1)}(u_1,u_3)\frac{1}{(u_1^+ u_2^+)}\Bigg)
\nonumber\\
&& - D^+_{1a} D^+_{1b}\,\widetilde q^+(q+p,\theta,u_3) \lambda(-p,\theta, u_1)\, q^+(-q,\theta,u_2)\,
\frac{(\xi_0-1)(\widetilde\gamma^M)^{ab} k_M}{2 k^4 (q+k)^2 (q+k+p)^2}\,(u_3^+ u_2^+)\nonumber\\
&&\times \Bigg(\frac{1}{(u_1^+ u_2^+)^2} D^{--}_3 \delta^{(0,0)}(u_1,u_3) + \frac{1}{(u_1^+ u_3^+)^2} D^{--}_2 \delta^{(0,0)}(u_1,u_2)\Bigg)
\Bigg]\Bigg\}. \label{Expr10}
\end{eqnarray}

\noindent
Then we integrate by parts with respect to the derivatives $D^{--}$  and take off one harmonic integral with the help of the delta functions.
Taking into account that the first term vanishes as a consequence of the analyticity of the superfields $\lambda$, $\widetilde q^+$, and $q^+$,
the expression (\ref{Expr10}) can be further rewritten as

\begin{eqnarray}
&&\hspace*{-8mm} 2f_0^2 \int \frac{d^{6}p}{(2\pi)^6}\,\frac{d^{6}q}{(2\pi)^6}\,\frac{d^{6}k}{(2\pi)^6}\, \frac{(\xi_0-1)}{k^4 (q+k)^2 (q+k+p)^2} \int d^8\theta\, du_1\,\lambda(-p,\theta, u_1) \Bigg\{\int du_2\, \frac{1}{(u_1^+ u_2^+)}\nonumber\\
&&\hspace*{-8mm}\times q^+(-q,\theta,u_2)\, \Bigg[ \frac{1}{2}(D^+_{1})^4\, (D_1^{--})^2 \widetilde q^+(q+p,\theta,u_1) - k^2 \widetilde q^+(q+p,\theta,u_1)
+ \frac{1}{2} (\widetilde\gamma^M)^{ab} k_M\, D^+_{1a} \nonumber\\
&&\hspace*{-8mm}\times\, D^+_{1b} D_1^{--}\,\widetilde q^+(q+p,\theta,u_1) \Bigg]
+ \int du_3\, \frac{1}{(u_1^+ u_3^+)} \Bigg[ \frac{1}{2} (D^+_{1})^4\,\widetilde q^+(q+p,\theta,u_3)\, (D_1^{--})^2 q^+(-q,\theta,u_1) \nonumber\\
&&\hspace*{-8mm} - k^2 \widetilde q^+(q+p,\theta,u_3)\, q^+(-q,\theta_1,u_1)
- \frac{1}{2} (\widetilde\gamma^M)^{ab} k_M\, D^+_{1a} D^+_{1b}\,\widetilde q^+(q+p,\theta,u_3)\,  D_1^{--} q^+(-q,\theta,u_1) \Bigg] \Bigg\}.\nonumber\\
\end{eqnarray}

\noindent
Once again, integrating by parts and taking into account that

\begin{equation}
\frac{1}{2} (D^{+})^4 (D^{--})^2 = \partial^2;\qquad (\widetilde\gamma^M)^{ab} D^+_{1a} D^+_{1b} D_1^{--} = - 4i\partial^M
\end{equation}

\noindent
on the analytic superfields, this expression can be  cast in the form

\begin{eqnarray}
&&\hspace*{-6mm} - 2f_0^2 \int \frac{d^6p}{(2\pi)^6}\, \frac{d^6q}{(2\pi)^6}\, \frac{d^6k}{(2\pi)^6} \frac{(\xi_0-1)}{k^4 (k+q)^2} \int d^8\theta\, du_1\, du_2\, \frac{1}{(u_1^+ u_2^+)}
\widetilde q^+(q+p,\theta, u_1) \lambda(-p,\theta,u_1) \qquad\nonumber\\
&&\hspace*{-6mm} \times  q^+(-q,\theta\,u_2) - 2f_0^2 \int \frac{d^6p}{(2\pi)^6}\, \frac{d^6q}{(2\pi)^6}\, \frac{d^6k}{(2\pi)^6} \frac{(\xi_0-1)}{k^4 (k+q+p)^2} \int d^8\theta\, du_1\, du_3\,  \frac{1}{(u_1^+ u_3^+)} \widetilde q^+(q+p,\theta, u_3) \nonumber\\
&&\hspace*{-6mm} \times  \lambda(-p,\theta,u_1) q^+(-q,\theta\,u_1),\vphantom{\frac{1}{2}}
\end{eqnarray}

\noindent
where we have also used the relations

\begin{eqnarray}
(q+p)^2 + k^2 + 2 k_M (q+p)^M = (q+k+p)^2\,, \qquad q^2 + k^2 + 2 k_M q^M = (q+k)^2.
\end{eqnarray}

\section{Gauge-dependent part of the effective action and the hypermultiplet equations of motion}
\hspace*{\parindent}\label{Appendix_Shell}

In this appendix we verify that the gauge-dependent part of the effective action vanishes on shell. This is an important non-trivial check of the  correctness of our calculations.

First, we consider the two-point Green function of the hypermultiplet  given by Eq. (\ref{Hypermultiplet_Green_Function}). Using the identity

\begin{equation}\label{Identity1}
\qquad\frac{1}{(u_1^+ u_2^+)} = D^{++}_1 \frac{(u_1^- u_2^+)}{(u_1^+ u_2^+)^2} + D^{--}_1 \delta^{(1,-1)}(u_1,u_2) = D^{++}_1 D^{++}_2 \frac{(u_1^- u_2^-)}{(u_1^+ u_2^+)^2} + D^{--}_1 \delta^{(1,-1)}(u_1,u_2),\qquad
\end{equation}

\noindent
we rewrite it as

\begin{eqnarray}
&& \widetilde\Gamma^{(2)} = - 2if_0^2 \int \frac{d^6p}{(2\pi)^6}\,\frac{d^6k}{(2\pi)^6} \frac{(\xi_0-1)}{k^4 (k+p)^2} \int d^8\theta\, du_1\, du_2\, \Big( D^{++}_1 D^{++}_2\frac{(u_1^- u_2^-)}{(u_1^+ u_2^+)^2} \qquad\nonumber\\
&& + D^{--}_1 \delta^{(1,-1)}(u_1,u_2) \Big) \widetilde q^+(p,\theta, u_1) q^+(-p,\theta\,u_2).\qquad
\end{eqnarray}

\noindent
The second term in this expression vanishes due to the analyticity of the hypermultiplet superfield,

\begin{equation}
\int d^8\theta\, du D^{--} \widetilde q^+(p,\theta,u) q^+(-p,\theta,u) =  \int d^4\theta^+\, du\, (D^+)^4 \Big(D^{--} \widetilde q^+(p,\theta,u) q^+(-p,\theta,u)\Big) = 0.
\end{equation}

\noindent
After integrating  by parts with respect to the harmonic derivatives, the considered contribution to the effective action can be represented as

\begin{equation}\label{Hypwermultiplet_Gauge_Depandence}
- 2if_0^2 \int \frac{d^6p}{(2\pi)^6}\, \frac{d^6k}{(2\pi)^6} \frac{(\xi_0-1)}{k^4 (k+p)^2}
\int d^8\theta\, du_1\, du_2\, \frac{(u_1^- u_2^-)}{(u_1^+ u_2^+)^2} D^{++} \widetilde q^+(p,\theta, u_1)\, D^{++} q^+(-p,\theta\,u_2).
\end{equation}

\noindent
Using the equations of motion for the hypermultiplets

\begin{equation}
0 = \nabla^{++} q^+ = \big(D^{++} + i V^{++}\big)q^+;\qquad 0 = \nabla^{++} \widetilde q^+ = \big(D^{++} - i V^{++}\big)\widetilde q^+,
\end{equation}

\noindent
we see that on shell the expression (\ref{Hypwermultiplet_Gauge_Depandence}) is proportional  to $\widetilde q^+ (V^{++})^2 q^+$.
However, in this paper we do not consider terms quadratic in the gauge superfield $V^{++}$. This implies that, within the accuracy of our approximation,
the part of the one-loop effective action corresponding to the hypermultiplet two-point function vanishes on shell.

Next, we consider the gauge dependent part of the three-point gauge superfield - hypermultiplet Green function. It corresponds
to the terms proportional to $\widetilde q^+ V^{++} q^+$ in the expression (\ref{Gamma_Gauge_Dependent}). We will demonstrate that $\widetilde \Gamma^{(3)}$
vanishes on shell (in the approximation  when all terms with more than one $V^{++}$ are omitted).

Using the identity

\begin{equation}\label{Identity2}
\frac{1}{(u_1^+ u_2^+)^2} = D_2^{++} \frac{(u_2^- u_1^+)}{(u_2^+ u_1^+)^3} + \frac{1}{2} (D_2^{--})^2 \delta^{(2,-2)}(u_2,u_1)
\end{equation}

\noindent
and discarding terms quadratic in $V^{++}$ (coming from $D^{++} q^+$ and $D^{++}\widetilde q^+$ after using the equations of motion), we obtain

\begin{eqnarray}\label{Term1}
&&\int d^8\theta\, du_1\, du_2\, \widetilde q^+(q+p,\theta,u_1) V^{++}(-p,\theta,u_2) q^+(-q,\theta,u_1) \frac{1}{(u_1^+ u_2^+)^2}\nonumber\\
&& \longrightarrow \frac{1}{2} \int d^8\theta\, du\, \widetilde q^+(q+p,\theta,u) (D^{--})^2 V^{++}(-p,\theta,u) q^+(-q,\theta,u) \nonumber\\
&& = \frac{1}{2} \int d^4\theta^+\, du\, \widetilde q^+(q+p,\theta,u) (D^+)^4 (D^{--})^2 V^{++}(-p,\theta,u) q^+(-q,\theta,u) \qquad\nonumber\\
&& = -p^2 \int d^4\theta^+\, du\, \widetilde q^+(q+p,\theta,u) V^{++}(-p,\theta,u) q^+(-q,\theta,u),
\end{eqnarray}

\noindent
where the arrow indicates that we omitted some terms vanishing on shell, as well as  $O((V^{++})^2)$ terms.

Using Eq. (\ref{Identity1}) twice, we have

\begin{eqnarray}
&& \int d^8\theta\, du_1\, du_2\, du_3\,\widetilde q^+(q+p,\theta,u_1) V^{++}(-p,\theta, u_2)\, q^+(-q,\theta,u_3) \frac{1}{(u_1^+ u_2^+) (u_2^+ u_3^+)}\qquad\nonumber\\
&&  \longrightarrow  - \int d^8\theta\, du\, D^{--}\widetilde q^+(q+p,\theta,u) V^{++}(-p,\theta,u) D^{--} q^+(-q,\theta,u) \nonumber\\
&& = - 2 q^M (q+p)_M \int d^4\theta^+\, du\, \widetilde q^+(q+p,\theta,u) V^{++}(-p,\theta,u) q^+(-q,\theta,u).
\end{eqnarray}

\noindent
The remaining terms vanish. Indeed, let us consider the expression

\begin{equation}
\int du_1\, du_2\, du_3\,  D^+_{2a} D^+_{2b}\,\widetilde q^+(q+p,\theta,u_1)  V^{++}(-p,\theta, u_2)\, q^+(-q,\theta,u_3)\,
\frac{(u_1^+ u_3^+)}{(u_1^+ u_2^+)^2 (u_2^+ u_3^+)^2} \nonumber\\
\end{equation}

\noindent
and make use of the relation $(u_1^+ u_3^+) = D^{++}_1 D^{++}_3 (u_1^- u_3^-)$. Then, after integrating by parts with respect to the harmonic derivatives
$D^{++}_1$ and $D^{++}_3$,
up to the terms quadratic in $V^{++}$,  we observe that on shell the resulting expression  is proportional to $(u_1^- u_1^-) = 0$,

\begin{eqnarray}
&& (92) \longrightarrow \int du_1\, du_2\, du_3\,  D^+_{2a} D^+_{2b}\,\widetilde q^+(q+p,\theta,u_1)  V^{++}(-p,\theta, u_2)\, q^+(-q,\theta,u_3)\,(u_1^- u_3^-)\nonumber\\
&&\times  D^{--}_1 \delta^{(2,-2)}(u_1,u_2) D^{--}_3 \delta^{(2,-2)}(u_3,u_2) = 0.\vphantom{\frac{1}{2}}
\end{eqnarray}

\noindent
Similarly, using the identity $(u_1^+ u_3^+)^2 = D^{++}_1 D^{++}_3\Big((u_1^- u_3^-)(u_1^+ u_3^+)\Big)$, we obtain

\begin{eqnarray}
&& \int du_1\, du_2\, du_3\,(D^+_{2})^4\,\widetilde q^+(q+p,\theta,u_1)  V^{++}(-p,\theta, u_2)\, q^+(-q,\theta,u_3)\,
\frac{(u_1^+ u_3^+)^2}{(u_1^+ u_2^+)^3 (u_2^+ u_3^+)^3} \nonumber\\
&& \longrightarrow \frac{1}{4} \int du_1\, du_2\, du_3\,(D^+_{2})^4\,\widetilde q^+(q+p,\theta,u_1)  V^{++}(-p,\theta, u_2)\, q^+(-q,\theta,u_3)\, (u_1^- u_3^-)(u_1^+ u_3^+)\qquad\nonumber\\
&& \times (D^{--}_1)^2 \delta^{(2,-2)}(u_1,u_2) (D^{--}_3)^2 \delta^{(2,-2)}(u_3,u_2) = 0.\vphantom{\frac{1}{2}}
\end{eqnarray}

Finally, collecting all terms, we conclude that the exploiting of the  hypermultiplet equations of motion allows us to rewrite the part of $\widetilde\Gamma$ corresponding
to the three-point gauge superfield - hypermultiplet Green function in the form

\begin{eqnarray}
&& \widetilde\Gamma^{(3)} = 2f_0^2 \int \frac{d^{6}p}{(2\pi)^6}\,\frac{d^{6}q}{(2\pi)^6}\,\frac{d^{6}k}{(2\pi)^6}\,
\frac{(\xi_0-1)}{k^2 (q+k)^2 (q+k+p)^2} \Big((q+p)^2 + q^2\Big)\nonumber\\
&&\times \int d^4\theta^+\, du\, \widetilde q^+(q+p,\theta,u) V^{++}(-p,\theta,u) q^+(-q,\theta,u).
\end{eqnarray}

\noindent
For the on-shell hypermultiplets the relations $q^2 =0$ and $(q+p)^2=0$ are valid, so this expression vanishes. The conclusion is that the gauge-dependent contributions
to the effective action are indeed canceled on shell in the approximation we stick to.

\end{document}